\documentclass[referee,sn-apa]{sn-jnl}

\usepackage{graphicx}
\usepackage{paralist}
\usepackage{multirow}%
\usepackage{amsmath,amssymb,amsfonts}%
\usepackage{amsthm}%
\usepackage{mathrsfs}%
\usepackage[title]{appendix}%
\usepackage{xcolor}%
\usepackage{textcomp}%
\usepackage{manyfoot}%
\usepackage{booktabs}%
\usepackage{algorithm}%
\usepackage{algorithmicx}%
\usepackage{algpseudocode}%
\usepackage{listings}
\usepackage{xcolor}
\usepackage{ amssymb }
\usepackage{nicefrac}
\usepackage{siunitx}
\usepackage{subcaption}

\newcommand{\Chaeyeon}[1]{\textcolor{black}{#1}}

\newcommand{\Subhro}[1]{\textcolor{black}{#1}}

\begin{document}

\title[Article Title]{Understanding Pedestrian Movement Using Urban Sensing Technologies: The Promise of Audio-based Sensors}

\author[1]{\fnm{Chaeyeon} \sur{Han}}\email{chan303@gatech.edu}

\author[2]{\fnm{Pavan} \sur{Seshadri}}\email{pseshadri9@gatech.edu}
\equalcont{These authors contributed equally to this work.}

\author[2]{\fnm{Yiwei} \sur{Ding}}\email{yding402@gatech.edu}
\equalcont{These authors contributed equally to this work.}

\author[3]{\fnm{Noah} \sur{Posner}}\email{noah.posner@ipat.gatech.edu}
\equalcont{These authors contributed equally to this work.}

\author[4]{\fnm{Bon Woo} \sur{Koo}}\email{bonwookoo@torontomu.ca}
\equalcont{These authors contributed equally to this work.}

\author[1]{\fnm{Animesh} \sur{Agrawal}}\email{anagrawal@gatech.edu}
\equalcont{These authors contributed equally to this work.}

\author[2]{\fnm{Alexander} \sur{Lerch}}\email{alexander.lerch@gatech.edu}
\equalcont{These authors contributed equally to this work.}

\author*[1]{\fnm{Subhrajit} \sur{Guhathakurta}}\email{subhro.guha@design.gatech.edu}
\equalcont{These authors contributed equally to this work.}

\affil*[1]{\orgdiv{Center for Spatial Planning Analytics and Visualization}, \orgname{Georgia Tech}, \orgaddress{\city{Atlanta}, \state{GA}, \country{USA}}}

\affil[2]{\orgdiv{Music Informatics Group}, \orgname{Georgia Tech}, \orgaddress{\city{Atlanta}, \state{GA}, \country{USA}}}

\affil[3]{\orgdiv{IPaT}, \orgname{Georgia Tech}, \orgaddress{\city{Atlanta}, \state{GA}, \country{USA}}}

\affil[4]{\orgdiv{School of Urban \& Regional Planning}, \orgname{Toronto Metropolitan University}, \orgaddress{\city{Toronto}, \state{ON}, \country{CANADA}}}


\abstract{While various sensors have been deployed to monitor vehicular flows, sensing pedestrian movement is still nascent. Yet walking is a significant mode of travel in many cities, especially those in Europe, Africa, and Asia. Understanding pedestrian volumes and flows is essential for designing safer and more attractive pedestrian infrastructure and for controlling periodic overcrowding. This study discusses a new approach to scale up urban sensing of people with the help of novel audio-based technology. It assesses the benefits and limitations of microphone-based sensors as compared to other forms of pedestrian sensing. A large-scale dataset called ASPED is presented, which includes high-quality audio recordings along with video recordings used for labeling the pedestrian count data. The baseline analyses highlight the promise of using audio sensors for pedestrian tracking, although algorithmic and technological improvements to make the sensors practically usable continue. This study also demonstrates how the data can be leveraged to predict pedestrian trajectories. Finally, it discusses the use cases and scenarios where audio-based pedestrian sensing can support better urban and transportation planning.}




\maketitle
\newpage
\section{Introduction}\label{sec:intro}
A significant component of smart city initiatives has been the deployment of sensor technologies to monitor and control various city services and functions. Cities use a variety of sensors to assess how urban services are being delivered and accessed, which helps alleviate bottlenecks and trigger advance warnings about potential service disruptions. Understanding the temporal and spatial variation in the demand for urban services can also lead to better resource use, more equitable service delivery, and greater sustainability and resilience. Various sensors are now deployed in the urban environment, particularly in transportation, but also to monitor environmental conditions, the flow of energy, water, and waste, and in tracking criminal activities \citep{lee_towards_2014}. More recently, with the growing interest in active mobility and walkability, several cities have experimented with various technologies to sense people. 

While the movement of vehicles has been an important component of traffic planning, there has been less effort in understanding the movement of pedestrians in cities, particularly in the United States. By understanding pedestrian flows, we can design better and more equitable walkable environments and gain useful insights into the spaces people linger in or avoid, as well as about the activities they pursue in the public domain. Accurate prediction of individual and social behavior in public spaces has powerful implications for urban planning. 

The detection of pedestrians has been mainly based on video data analysis \citep{rahman2019real, li2016pedestrian} or through infrared counters \citep{yang2010investigating, yang2011enhancing, mathews2009evaluation}; both are several times more expensive than audio sensing. More sophisticated alternatives that are sometimes considered for pedestrian sensing, such as radar, radio beams, inductive loops, and piezoelectric strips, are also costly to deploy and maintain \citep{ozan2021state, ozbay_automated_2010}. In this paper, we explore the potential for microphone-based sensors, combined with methods developed for the analysis of highly complex musical audio signals, to be adapted to sensing pedestrians. As sensors, microphones offer several advantages: they are affordable, have low power requirements, can cover large angles up to 360 degrees, and can capture otherwise unobserved data given that sound travels around objects and thus cannot be easily blocked. \Subhro{Recent research has also demonstrated that soundscapes detected through audio devices provide sufficient visual information of the same places \citep{ZHUANG2024102122}.} These advantages are counterbalanced by challenges that require investigation, experimentation, and mitigation, including the requirement to develop more advanced processing algorithms for extracting meaningful information as multiple sound signals are superposed in a poly-timbral mixture with unknown directional information; the challenge of positioning microphone-based sensors in ways that optimize data collection; and challenges beyond technology such as issues of privacy, anonymity, and sanitization of data. 

Given that prior research has provided little guidance on how to mitigate many of these challenges, we have addressed this gap by conducting pilot experiments in a campus setting to identify an appropriate and sustainable way of achieving continuous data capture and processing. In this paper, we report on our preliminary tests of using microphones to sense people while also demonstrating how strategically placed sensors can offer valuable information about pedestrian flows. We demonstrate our approach by using audio data to detect pedestrians. However, the \Chaeyeon{flow} prediction algorithm uses the information we retrieved from the video footage used to label the audio files. We use these separate approaches because our algorithms to detect pedestrians using audio are not at the level of accuracy that video-based methods have reached. However, the \Chaeyeon{flow} prediction algorithms we develop are agnostic to the method of obtaining pedestrian counts. We expect that audio-based pedestrian count data with improved accuracy can easily substitute for the video-based data we use now for pedestrian \Chaeyeon{flow} tracking.

The rest of the paper is organized as follows: the next section, Section~\ref{sec:sensing}, assesses current pedestrian sensing technologies and provides a brief overview of their technical challenges. Section~\ref{sec:tracking} reviews the pedestrian flow \Chaeyeon{detecting }approaches that inform our study. Next, in Section~\ref{sec:research}, we discuss the collection and curation of the ASPED dataset, which was used in our preliminary analyses. In this section, we also present our methodological approach and the design of experiments. Section~\ref{sec:results} presents our results from the experiments in pedestrian detection and for predicting pedestrian \Chaeyeon{flows}. Section~\ref{sec:discussion} offers a discussion of the promise and broader impacts of audio-sensing technology for sensing people and concludes with some closing remarks.

\section{Pedestrian sensing technologies}\label{sec:sensing}
While pedestrian sensors have been around for over 30 years, audio-based sensing has ---~to the best knowledge of the authors~--- never been attempted. This is probably because data science and machine learning-based technologies were less mature and are only now starting to transform many domains. Several pedestrian sensor systems have been operating in Europe and Australia since the early 1990s, and systematic efforts to deploy such systems in the U.S.\ began soon after 2000. The Federal Highway Administration commissioned a study in 2001 to evaluate whether automated pedestrian detectors at intersections can reduce pedestrian-vehicle conflicts when compared with standard pedestrian push-button walk signals \citep{hughes_evaluation_2001}. This study also compared a number of different pedestrian detection technologies, including microwave sensors, infrared detectors, and video. The detection technologies had a significant failure rate, and the authors concluded that ``improvements are needed in detection accuracy to reduce the number of false alarms and missed calls at intersections" \citep[p.16]{hughes_evaluation_2001}. Subsequently, the National Bicycle and Pedestrian Documentation Project (NBPDP) began in 2004 to standardize data collection of bicycles and pedestrians, initially using manual short-duration counts \citep{ozan_state---art_2021}. Soon, multiple federally funded studies were commissioned to evaluate various technologies to detect pedestrians in different locations, followed by similar efforts by many state and metropolitan government agencies \citep{ryus_methods_2014,minge_bicycle_2017,figliozzi_design_2014, fields_active_2012}. An excellent review of the reports and studies to date for bicycle and pedestrian data collection efforts and an evaluation of various sensor technologies to capture such data can be found in Ozan et al.~\citep{ozan_state---art_2021}.

Most past deployments of sensors to detect active mobility cannot distinguish pedestrians and bicyclists~\citep{ozan_state---art_2021}. The technologies used in these sensors, such as active and passive infrared, laser scanning, radar, and radio beams, are designed to detect interrupted or reflected pulses from human bodies or their heat signatures. Therefore, exposed humans, whether pedestrians, bicyclists, or people using micromobility options, will all be detected without distinction. Other technologies typically used in detecting vehicles, such as pneumatic tubes, inductive loops, and magnetometers are not appropriate for pedestrian tracking but can be used for sensing bicyclists. Other less common sensing technologies, such as ultrasonic sensors, pressure sensing mats, piezoelectric strips, and various hybrid technologies, have also been deployed for pedestrian counts. Each of these sensors offers particular advantages but also comes with several challenges and limitations. 

In the commercial realm, several companies have offered their own solutions to communities for tracking active mobility trips. Among the commercial offerings, Eco-Counter, MetroCount, Jamar, and StreetLight are some of the more popular vendors of bicycle and pedestrian data. Eco-Counter, a French company with offices in North America and across Europe, utilizes various technologies for its pedestrian and bicycle counters, including Pneumatic tubes, passive infrared, inductive loops, and mixed infrared/ inductive loops. Metrocount has developed bicycle counters (RidePod\circledR BT) using pneumatic tubes and pedestrian and bicycle counters that utilize piezoelectric technologies (RidepPod\circledR BP). More recently, companies such as Miovision and Numina have been offering video image-based sensors. These image-based sensors use advanced neural networks to detect and isolate different transport and pedestrian traffic modes. Perhaps the most advanced private player in this domain is StreetLight Data, which is pioneering the development of proprietary \Chaeyeon{artificial intelligence (AI)} algorithms applied to data from millions of mobile devices, \Chaeyeon{Internet of Things (IoT)} sensors, and other geospatial databases to estimate traffic mode and volumes at a fine geographic scale. Regardless of the technology and their providers, all the technologies noted above have a price point that is well above \$1,500 per sensor, making them cost-prohibitive for wide-scale adoption across urban areas. Therefore, they are mostly used in a few selected locations and are applied to track active mobility in small areas within the city.

At this time, the most widely used pedestrian sensing technologies are:
\begin{inparaenum}[(i)]
    \item   video image processing, and
    \item   active and passive infrared counters. 
\end{inparaenum}

\subsection{Video-based sensing using computer vision}
In the absence of advanced machine learning approaches and training data sets, video technology was initially used to count pedestrians manually from recorded video \citep{ozan_state---art_2021}. With the rapid progress in computer vision-based algorithms and growing availability of pre-tagged image datasets in the last two decades, detection of people and objects from video or static images has become easier and more efficient \citep{dollar_pedestrian_2009,martin_learning_2004,barron_performance_1994,baker_database_2011,scharstein_taxonomy_2002,fei-fei_one-shot_2006}. This research area has seen rapid development also due to its importance for autonomous, self-driving vehicles. In a review article from 2012, Dollar et al.\ point out that a performance drop is noticed across various systems if pedestrians are represented by less than 80 pixels \citep{dollar_pedestrian_2012}. Even in the best case, however, up to 20 percent of pedestrians can be missed. The data quality strongly impacts the miss rate, as can be seen through strong variance over different datasets \citep{brunetti_computer_2018} and weather conditions \citep{li_deep_2020}.

It should be noted, however, that most of the work studying pedestrian detection aims at solving the detection problem under the very challenging circumstances of an autonomous vehicle, where the video camera itself is moving, the pedestrians not only have to be counted but individually tracked, the video feed potentially has limited resolution, and the error of missing a pedestrian can have fatal consequences. Our objective is different and can tolerate a somewhat higher margin of error in prediction.  

\subsection{Infrared counters}
Three types of infrared sensors have been used for pedestrian detection: 
\begin{inparaenum}[(i)]
    \item   active,
    \item   passive, and 
    \item   target reflective. 
\end{inparaenum}
An active counter uses an invisible beam, which, when interrupted by a body, registers a pedestrian count. A passive infrared counter senses the heat emitted by human bodies passing through the sensing area to detect pedestrians. Target-reflective devices also use an invisible beam that is bounced back from a reflector mounted on the opposite side of a sensing area. The absence of the reflected beam indicates the presence of pedestrians. 

Several studies evaluated the performance of infrared counters and found them to be inaccurate, mostly because they systematically undercounted pedestrians. A study of trail users in Indiana that used infrared counters concluded that the sensor systematically undercounted trail users by 15\% \citep{wolter_summary_2001}. Other studies have also found similar results, such as a study conducted in San Diego County that found the undercount rate to be 15\% to 21\% from active infrared counters and 12\% to 48\% from passive infrared counters \citep{jones_seamless_2010}. The latest tests conducted in New Jersey found that the infrared undercounting error can be more than 20\% at sites with high volumes \citep{yang_investigating_2010, yang_enhancing_2011}. These field tests confirm that multiple pedestrians at the same time can confuse the infrared counters, especially when the bodies are lined up along the beam.

\section{Advances in pedestrian flow prediction}\label{sec:tracking}

The data gathered on pedestrian movement has proven invaluable for predicting and modeling human mobility patterns on an urban scale \citep{yabe_metropolitan_2023}. Predicting pedestrian flow plays a significant role in various fields, such as analyzing transportation and activity behaviors \citep{jiang_activity-based_2017, gonzalez_understanding_2008, ai_deep_2019}, managing disaster responses \citep{yabe_metropolitan_2023, oliver_mobile_2020}, assessing public safety, and urban planning \citep{ratti_mobile_2006}.
	
Recent advancements in pedestrian flow prediction show a growing inclination towards using \Chaeyeon{AI} \citep{kitano_od-network-based_2019}, particularly for handling the complexities of spatial-temporal prediction. The emergence of deep learning techniques, particularly the integration of Convolutional Neural Networks (CNNs) for spatial analysis and Recurrent Neural Networks (RNNs) for temporal patterns, has significantly improved the accuracy of these predictions \citep{ai_deep_2019}. Also, Deo and Trivedi \citep{deo_trajectory_2021} introduced a distinctive approach for grid-based prediction, a critical aspect of pedestrian and vehicle trajectory forecasting. They proposed multimodal trajectory forecasts on scenarios sampled from a grid-based policy, which is learned using maximum entropy inverse reinforcement learning (MaxEnt IRL). 

For regions with irregular spatial correlations, Graph Convolutional Networks (GCNs) have been heralded for their ability to model complex interactions between nodes effectively \citep{liu_pedestrian_2021, xia_3dgcn_2021}. These networks, where nodes represent areas and edges signify road links or Origin-Destination trajectories, have effectively processed complex spatial interactions. Liu et al.~\citep{liu_pedestrian_2021} utilized camera-detected pedestrian data and employed their unique GCN model, which is capable of directly processing the spatial topology of road networks. Xia et al.~\citep{xia_3dgcn_2021} introduced a 3-dimensional GCN (3DGCN) model for dynamic spatial-temporal graph prediction challenges, integrating Point-of-Interest (POI) data for enhanced accuracy. Furthermore, Sun et al.~\citep{sun_predicting_2022} proposed a hybrid of GCN and fully-connected neural networks, with a multi-view fusion module, to capture and predict spatial correlations and crowd flow dynamics. Studies have shown that these methods adeptly handle non-linear spatial dependencies and time-varying trends often present in pedestrian movements.

The inclusion of external factors, such as inter-region traffic and weather conditions, has also enriched pedestrian flow predictions. For instance, Zhang and Kabuka~\citep{zhang_combining_2018} and Zhang et al.~\citep{zhang_deep_2017} considered weather and day-specific influences, while Lin et al.~\citep{lin_deepstn_2019} utilized the DeepSTN+ model, integrating POIs and temporal elements for better spatial dependency understanding. Zhang et al.~\citep{zhang_predicting_2018} developed the ST-ResNet model to predict crowd inflow and outflow, incorporating weather and time data.

The studies mentioned previously have employed various data sources to enhance pedestrian flow predictions, including mobile \Chaeyeon{GPS (Global Positioning System)}, social media~\citep{zhang_deep_2017}, as well as data from bicycle sharing systems \citep{lin_deepstn_2019} and taxi GPS records \citep{sun_predicting_2022}. However, accessing this raw data for effective use in pedestrian mobility prediction poses a significant challenge for researchers. Often, simply having data is not sufficient; there is also a need for developing specialized software and systems, typically in collaboration with cellphone companies \citep{ratti_mobile_2006}. In this context, our audio-based pedestrian count dataset offers a potential solution to these challenges, providing a unique and valuable open resource for pedestrian mobility research.

\section{Data and Methods}\label{sec:research} 
\subsection{Data}
\subsubsection{Data collection and processing}
\begin{figure}
    \centering
    \includegraphics[width=\linewidth]{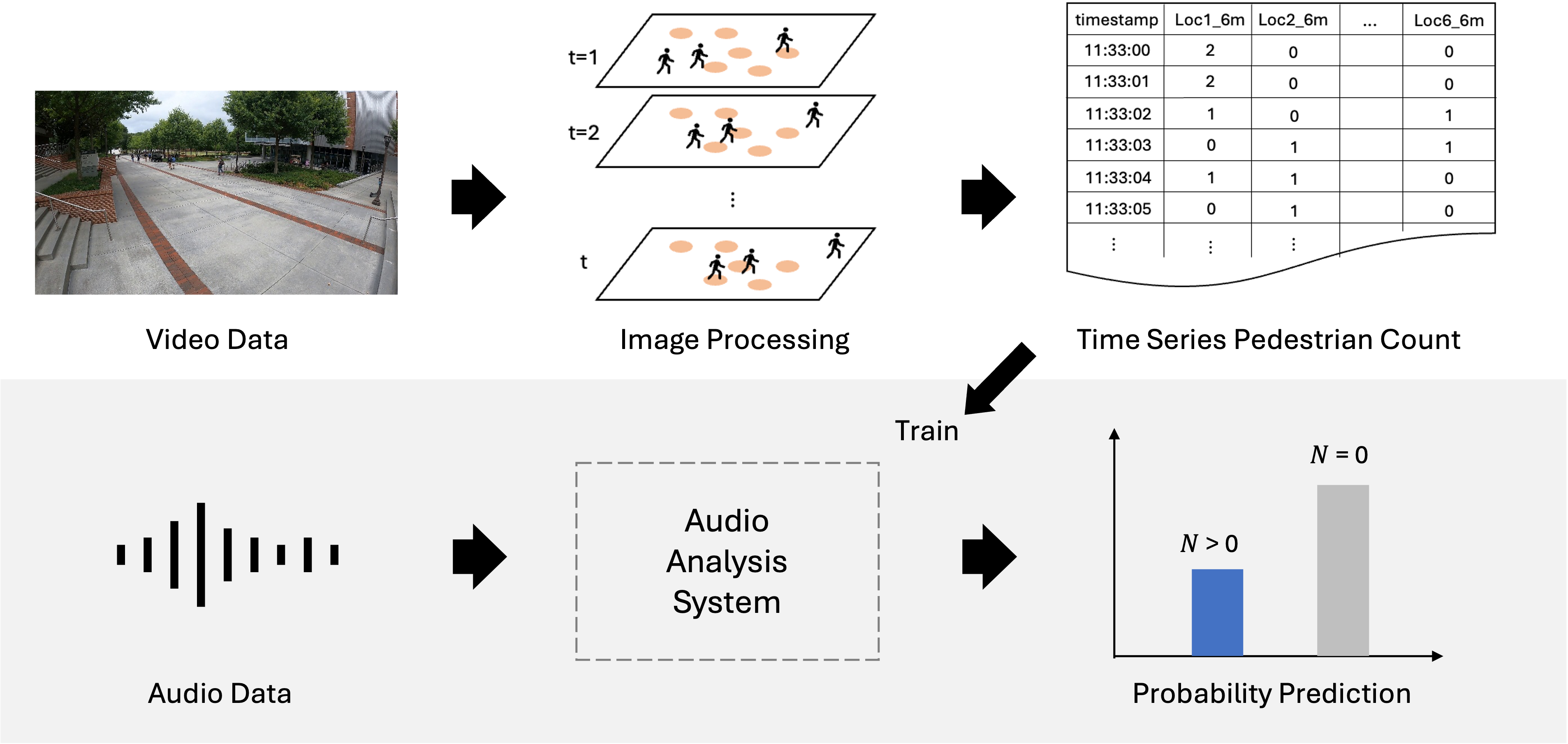}
    \caption{The overall workflow of training an audio analysis system to detect pedestrians.}
    \parbox{\linewidth}{\footnotesize Note: $N$ refers to the number of pedestrian detected.}
    \label{fig:flowchart}
\end{figure}

To evaluate the capability of microphone-based sensors in identifying pedestrian presence, we collected and analyzed a unique dataset, named Audio Sensing for PEdestrian Detection (ASPED) \citep{seshadri_asped_2024}.\footnote{\href{https://urbanaudiosensing.github.io/ASPED.html}{urbanaudiosensing.github.io/ASPED.html}, last access date Jan 31, 2024} This dataset serves as the foundation for a series of experiments that explore the possibility of audio sensing for pedestrian detection. The overall workflow is illustrated in Figure \ref{fig:flowchart}.

To facilitate data collection, we captured data using off-the-shelf recording devices. The audio collection setup used Tascam DR-05X recorders with power banks for extended recording duration, Saramonic SR-XM1 microphones to avoid RF interference issues of the Tascam's built-in mics, and 5L OverBoard Dry Flat Bags for weatherproofing while maintaining audio permeability. 
For video data, we used GoPro HERO9 Black cameras with USB pass-through doors connected to Anker PowerCore III Elite 26K power banks for longer recording. The power banks were enclosed in Seahorse 56 OEM Micro Hard Cases, modified with a drilled hole, a Wraparound Plastic Submersible Cord Grip for the cord, and a 90-degree USB connector for better positioning and fit. For synchronizing time across cameras, we displayed the time from \href{www.time.gov}{www.time.gov} on a mobile device to each camera after the recording started, followed by a whistle blow to mark the exact time, aiding in syncing with the audio recorders. In larger areas, multiple whistle signals were used.

The recording devices were set up in two locations on the Georgia Tech campus: \textit{Cadell Courtyard} and \textit{Tech Walkway}, both situated near dining areas but closed to vehicles. Due to the battery life of the devices, recording sessions were limited to approximately two days each. 

To extract the pedestrian count per video frame, we use the Mask2Former model \citep{Cheng2022} to detect people per frame at 1 frame per second. We use the specific implementation by OpenMMLab\footnote{\href{https://openmmlab.com}{openmmlab.com}, last access date Sep 5, 2023} which was trained on the Microsoft COCO dataset. This algorithm was parametrized with a prediction threshold of 0.7. For each frame of the video, the algorithm identified the `person' class and generated bounding boxes around them. Subsequently, to analyze the proximity of these detected pedestrians to the audio recorders, circular buffers with various radii $r \in [\SI{1}{\meter}, \SI{3}{\meter}, \SI{6}{\meter}, \SI{9}{\meter}]$ were superimposed on the video frames. These buffers were centered around the poles where the audio recorders were mounted. The orientation of the buffers was adjusted to align with the perspective of each specific video recording. Finally, every frame was annotated with the number of pedestrians detected if the bottom-center point of pedestrian bounding boxes intersected with the recorder buffers; otherwise, it was labeled as no-pedestrian present.

Note that the experiment presented in the following only utilizes binary labels (pedestrian present vs.\ no pedestrian present or pedestrian count  $N \geq N_\mathrm{Threshold}$ vs.\ count $N<N_\mathrm{Threshold}$), however, the actual dataset labels reflect the actual pedestrian count.

\subsubsection{Data description}

Overall, we captured one frame-per-second video recording totaling 3,406,229 video frames, accompanied by nearly 2,600 hours of audio, in five recording sessions. All recording days were weekdays with only one exception.

A noteworthy characteristic of the dataset is its imbalance: most of the time, there is no pedestrian close to the microphones. Therefore, the number of frames annotated with a pedestrian count of zero is considerably higher than those with non-zero pedestrians.
Across the five collected sessions, the percentage of frames with one or more pedestrians ranges from $4.58\%$ to $10.75\%$ with a mean of $8.79\%$.

\begin{figure}
    \centering
    \includegraphics[width=\linewidth]{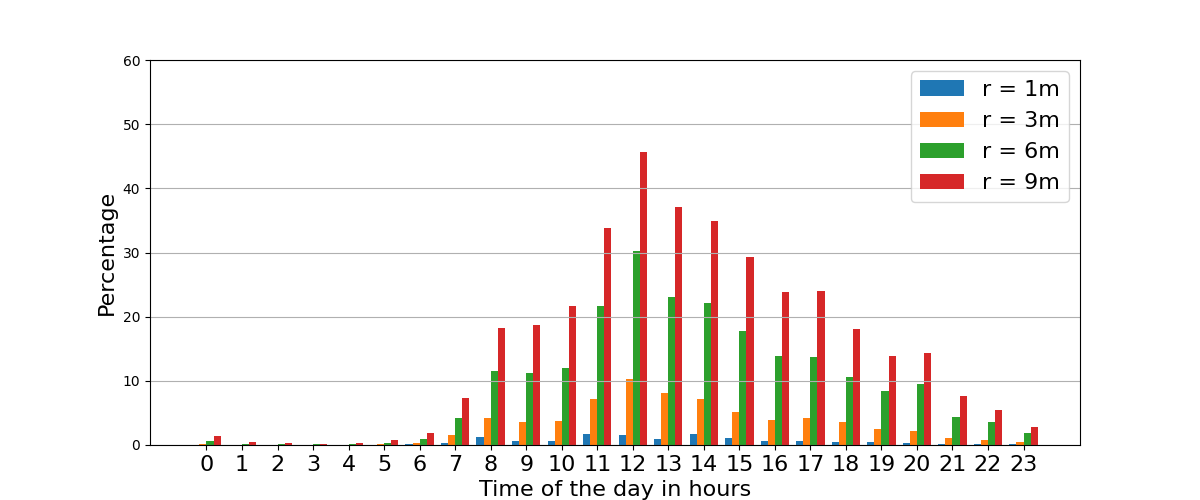}
    \caption{Percentage of labels with pedestrians in different hours during a day.}
    \label{fig:percentage_by_hour}
\end{figure}

    By looking at the distribution over different hours in a day in Figure ~\ref{fig:percentage_by_hour}, we observe that most pedestrian events happen~---unsurprisingly---~during the daytime and that there is a small peak during lunchtime around 12 PM when $20\%$ to $30\%$ of the frames have pedestrians in proximity to the microphone. Between 1 AM and 4 AM, on the other hand, there are hardly any frames with pedestrians.

This label imbalance is important to keep in mind as it requires a careful design of machine learning models, training strategies, and evaluation methodology. Otherwise, the model might easily achieve (seemingly) high accuracy by simply predicting everything as no-pedestrian without learning anything meaningful. 

We conducted experiments using data from audio and video sensors to achieve two objectives: 
\begin{inparaenum}[(i)]
\item detect the presence of pedestrians near audio sensors using audio data (specifically when $N\neq 0$) that is annotated with the help of video frames from the same place and time, and 
\item predict the count of pedestrians near each sensor based on information from video frames in the ensuing seconds.
\end{inparaenum}
In the subsequent sections, we explain our methodological approach for addressing each objective.

\subsection{Audio Sensing for Pedestrian Detection}

\subsubsection{Models}
As mentioned above, the recorded audio signals are a superposition of many source signals comprising an auditory scene, which may or may not indicate pedestrians. Sounds indicating pedestrians can include, e.g., footsteps or speech, while other sounds can originate from various sources, including traffic, animals, construction, etc. All of these sounds are mixed together in a time-variant superposition with different volumes. This complexity requires sophisticated audio analyses such as the ones used in the field of musical audio analysis~\citep{lerch_introduction_2023}. In this field, traditional approaches based on audio descriptors extracted from the time or spectral domain of the signal with a subsequent classifier (e.g., Support Vector Machine) have been nearly completely replaced by deep learning approaches as such approaches have shown superior performance across a wide variety of tasks, particularly challenging tasks with complex mixtures of audio sources.

\begin{figure}
    \centering
    \includegraphics[width=\linewidth]{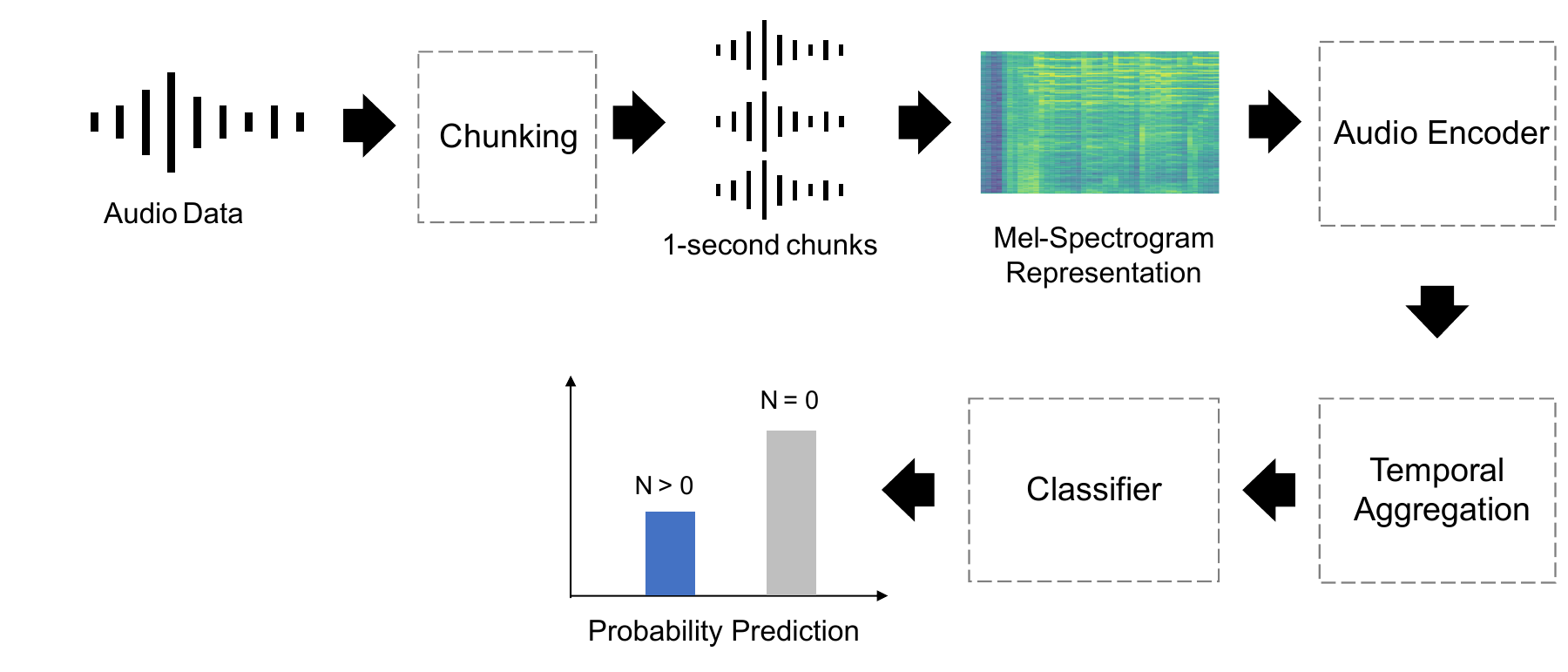}
    \caption{The overall workflow of our audio analysis system.}
    \label{fig:aas}
    \parbox{\linewidth}{\footnotesize Note: $N$ refers to the number of pedestrian detected.}
\end{figure}

A more detailed schematic of the ``Audio Analysis System'' introduced in Figure~\ref{fig:flowchart} is presented in Figure~\ref{fig:aas}. Essentially, a binary classifier is trained which predicts whether a pedestrian is recorded in a given audio sample. Here, a $t$-second audio recording is chunked into individual \SI{1}{s} segments, transformed into a Mel Spectrogram, and fed into an audio encoder network that learns to extract the information relevant for pedestrian detection. The result is a 128-dimensional vector representing each second of audio. This vector is then fed into a temporal aggregation model to add contextual information about the activity across the entire audio sample. The output of the temporal aggregation is then the input to a classifier estimating the likelihood of pedestrians for each segment of audio. 

We investigated the following models to serve as the ``Audio Encoder'' shown in Figure~\ref{fig:aas}. These are methods for audio event classification at varying levels of complexity. 

\begin{enumerate}
    \item VGGish Pre-Trained Features (0 learnable parameters)
    \item Convolutional Neural Network (2M learnable parameters)
    \item Audio Spectrogram Transformer (80M learnable parameters)
\end{enumerate}

The first model, denoted as VGGISH, uses a pre-trained VGGish network to extract audio information. The VGGish network \citep{hershey_cnn_2017} is originally trained using AudioSet, a dataset of short audio events used for audio classification tasks \citep{audioset}. This network is not updated during our training process. The second model, denoted as CONV, uses a six-layer convolutional network, trained from scratch using our data. The third model, denoted as AST, uses the Audio Spectrogram Transformer, which is a current SoTA model for audio classification tasks \citep{AST}. This model does not use the temporal aggregation model but produces per-second probabilities without longer context. For the temporal aggregation for VGGISH and CONV, we use a transformer encoder, a SoTA model used for sequence modeling \citep{transformer}. All models are trained using a binary cross entropy loss for binary classification. For details regarding hyperparameters and the implementation of our models, please see our ASPED dataset publication \citep{seshadri_asped_2024}. 

\subsubsection{Experimental setup}
By undertaking three experiments, we evaluated several attributes within our workflow, including model type, detection boundary, and strength of pedestrian signals.

 \textbf{Experiment~1 --- Comparison of performance of each audio encoder model}: We trained each aforementioned model in identical routines to determine relative performance.
 
 \textbf{Experiment~2 --- Comparison of performance across buffers of different radii}: We investigated the effect of the recording radius on the performance of each model by training and testing each model for different buffer radii $r \in [\SI{1}{\meter}, \SI{3}{\meter}, \SI{6}{\meter}, \SI{9}{\meter}]$. We anticipate that smaller radii have less but stronger signals of pedestrians, while larger radii have more but weaker signals of pedestrians. 
 
 \textbf{Experiment~3 --- Impact of training/testing thresholds on performance}:  To investigate the model's response to the strength of pedestrian signals, we set thresholds during training and testing for binary classification (i.e., only considering pedestrian counts above $N_\mathrm{Threshold}$-pedestrians as pedestrian events). We expect that pedestrian events under high thresholds correlate to stronger signals, while those under low thresholds correlate to weaker signals. In this study, we investigated setting $N_\mathrm{Threshold} \in [1,2,3,4]$ during both model training and testing.

 Due to the considerable class imbalance in our dataset, a classifier trained in a standard approach would likely learn to simply ignore the underrepresented class and predict the overrepresented class. To mitigate this, we add two enhancements to our training routine. First, during each training batch, we oversample the underrepresented class with replacement, such that each batch contains roughly half audio samples with pedestrian events. This exposes the model to each class roughly equally per-batch, with complete exposure to the dataset happening over multiple epochs of training, rather than per each epoch. Second, we apply a weighting function to our loss term, which roughly weights pedestrian events and no-pedestrian events equally in the loss term, such that both contribute to model learning equally. The function and overall loss term are shown below:
 \begin{equation}
        \mathcal{L} = \lambda\mathcal{L}_\mathrm{BCE+} + (1-\lambda)\mathcal{L}_\mathrm{BCE-}
    \end{equation}
    \begin{equation}
    \lambda = \left\{
    \begin{array}{lr}
        \frac{\nicefrac{1}{num^{+}}}{\nicefrac{1}{num^{+}} + \nicefrac{1}{num^{-}}}, & \text{if } num^{+} \neq 0\\
        0, & \text{if } num^{+} = 0
    \end{array}\right.    
\end{equation}


\subsection{Pedestrian Flow Prediction}
    \subsubsection{Experimental Setup}\label{sec:flow}
We also conducted a pilot study on a street level at a location referred to as \textit{Cadell Courtyard} within our dataset. Our ultimate goal is to forecast the inflow and outflow of pedestrians across street networks. However, this paper presents an initial step, where we concentrate on short-term predictions of the distribution of pedestrians across our sensors over time. Additionally, we propose a framework describing how our dataset could be improved and leveraged to conduct inflow-outflow prediction.

To achieve short-term pedestrian flow prediction, we undertake the following tasks:
\begin{inparaenum}[(i)]
    \item pedestrian detection from video sensors, 
    \item training a CNN using this data, and 
    \item utilizing the sliding window method for short-term prediction. 
\end{inparaenum}
Earlier sections of this paper have outlined the experimentation of the first step, and we are actively working on the subsequent stages as described in this section. 

\begin{figure}
    \includegraphics[width=\linewidth]{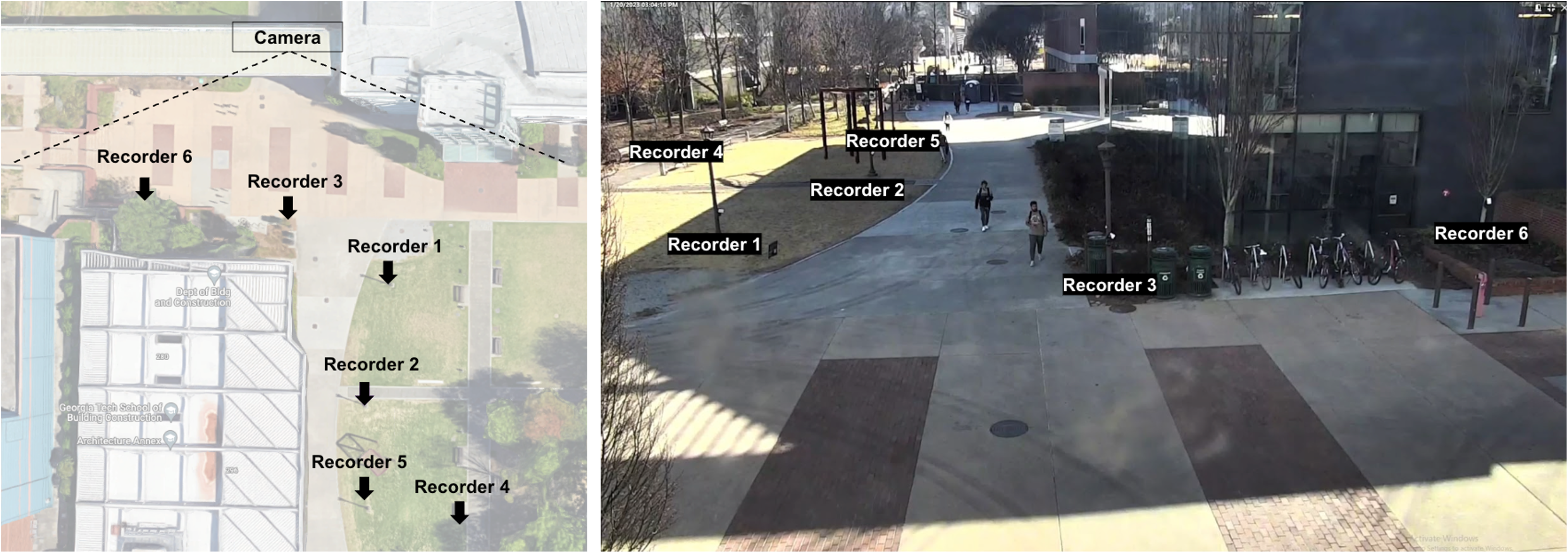}
    \caption{The Aerial (left) and the Surveillance Camera (right) View of the Experiment Site.}
    \label{fig:experiment-site}
\end{figure}

At the experiment site, we set up six audio recorders and employed a surveillance camera overseeing the entire area (Figure \ref{fig:experiment-site}). Recognizing that our methods for audio-based pedestrian sensing are in the process of refinement, we relied on the video feeds that are used to annotate the audio data. The intention is to develop pedestrian flow estimation algorithms with data extracted from video feeds and then update the validated models with audio-based sensor data when such data are deemed robust. Since our audio data does not contain information about pedestrian direction, we extracted only the count of pedestrians from video frames despite the capability to predict pedestrian directions. 

While simple mathematical methods like linear regression can be effective for such small-scale, densely arranged sensor networks, we encountered several limitations with this approach. For instance, linear regression requires creating separate models for each recorder, a process that can be inefficient for larger-scale predictions. Additionally, this method fails to account for the fluctuations in pedestrian flow that occur at different times of the day or during various seasons.

\begin{figure}
    \includegraphics[width=\linewidth]{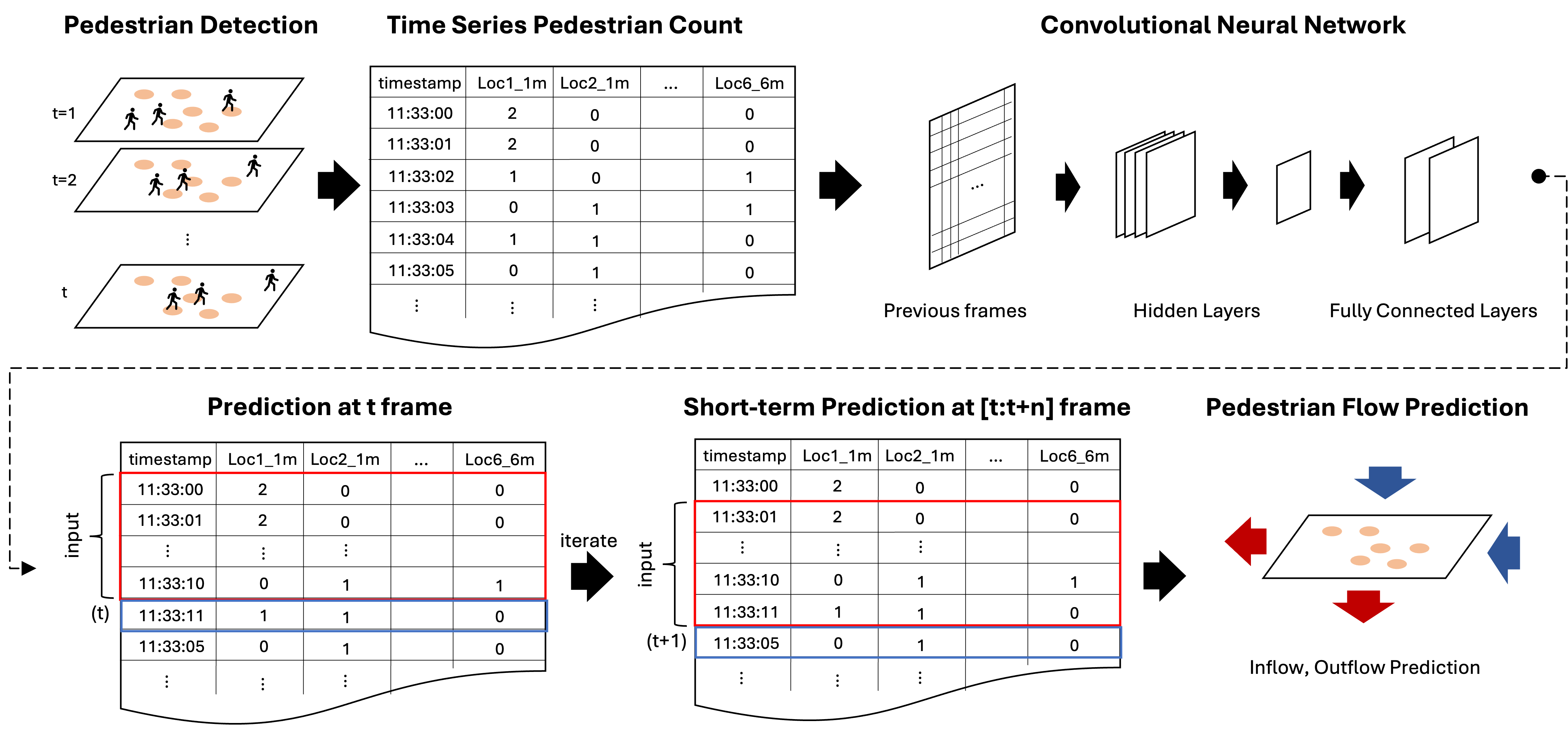}
    \caption{Pedestrian flow prediction framework.}
    \label{fig:framework}
\end{figure}

To address these issues and to better represent the nonlinear relationships between recorders, we formulated a predictive framework for estimating pedestrian flow utilizing a CNN (Figure~\ref{fig:framework}). This method processes the data of pedestrians detected in the past 11 frames (i.e., \SI{11}{\second}) from all recorders, to simultaneously predict pedestrian counts at each recorder location and for every radial distance (\SI{1}{\meter}, \SI{3}{\meter}, \SI{6}{\meter}, and \SI{9}{\meter}). The input for this CNN includes a comprehensive set of 25 data columns from all recorder locations and radii (6 locations times 4 boundaries), along with the timestamp, which could provide valuable insights into the time of day or date, potentially influencing pedestrian traffic flow patterns. 

Before feeding the data into the input layer, timestamps were converted to Unix timestamps for easier processing. Additionally, we excluded periods where all recorders consistently reported zero pedestrians over 11 consecutive frames, as these provide no useful data for prediction. Including these frames, our test accuracy would always be close to perfect, as over 90\% of our initial dataset has no pedestrians. We combined the five sessions in our dataset, which we randomly divided into training and testing sets using an 80/20 split. This approach allows for extensive training data and covers a wide range of times when pedestrian activity occurs. To standardize the data, we centered it around zero by subtracting the mean and scaled it by dividing it by the standard deviation, calculated element-wise for each input feature. These statistics were also applied to pre-process the test set data.

For CNN, we used a basic shallow architecture with 4 \Chaeyeon{2-dimensional} convolutional layers, 1 max pool layer, and 2 fully connected layers. We used a \Chaeyeon{stochastic gradient descent (SGD)} optimizer for 25 epochs with a learning rate of 0.001. 

\section{Results} \label{sec:results}
\subsection{Pedestrian Detection}
\begin{figure} 
    \begin{subfigure}[b]{0.5\linewidth}
    \centering
        \includegraphics[width=\linewidth]{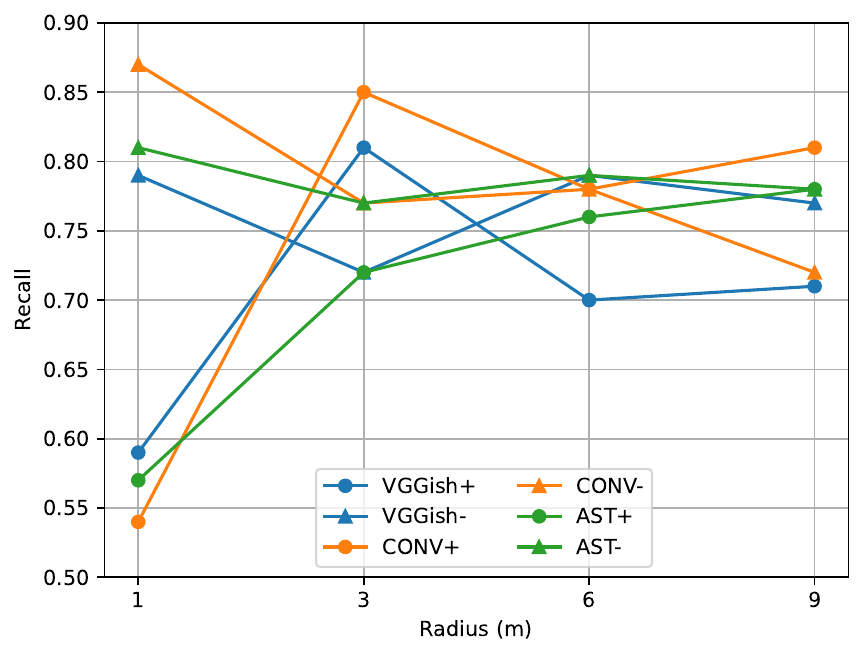}
        \caption{ }
        \label{fig:line}
    \end{subfigure}
    \begin{subfigure}[b]{0.5\linewidth}
        \centering
        \includegraphics[width=\linewidth]{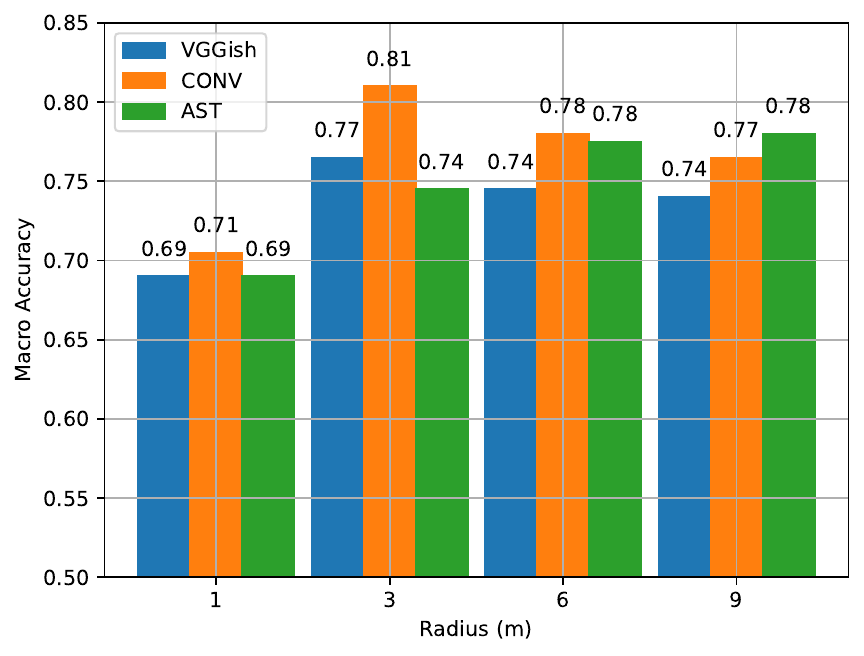}
        \caption{ }
        \label{fig:bar}
    \end{subfigure}
    \caption{(a)Recall for each class over each model and recording radius. Positive and negative classes are denoted by ”+” and ”-”,
        respectively.  (b)Macro average accuracy using the VGGISH, CONV, and AST models.}
    \parbox{\linewidth}{\footnotesize Source: \citep{seshadri_asped_2024}}
    \label{fig:results}
\end{figure}

Figures~\ref{fig:line} and \ref{fig:bar} present the model and radius comparisons for Experiments 1 and 2. We make the following observations. Generally, we find that audio encoders trained on our source task (CONV, AST) generally outperform the model not fine-tuned to pedestrian detection (VGGISH). CONV achieves the highest performance with middle radii 3 and 6, while VGGISH and AST are relatively constant from radii 3 to 9.  We also observe that negative class recall slightly outperforms positive class recall, as expected, due to data distributions. The mitigation by loss weighting and oversampling seems to work, however, as the effect is considerably less dramatic than the dataset imbalance itself. 

We note that while the test sets for each radius are identical in audio content, they are not identical in labels. Thus, the class distributions differ, and the results are not directly comparable. We find the most balanced performance using middle radii 3 and \SI{6}{\meter}, while radii 1 and \SI{9}{\meter} see a slight decline. Proximity to the microphone and shifting data distributions (higher count of pedestrian events as the radius increases) likely explain the change in performance among different radii.

\begin{figure}
    \centering
    \includegraphics[width=0.7\linewidth]{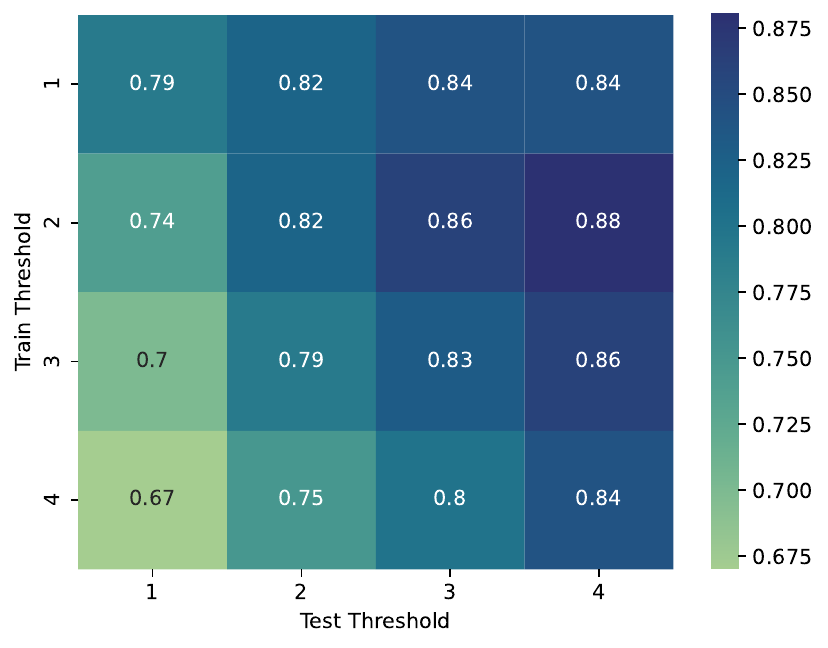}
    \caption{Macro average accuracy over each train and test pedestrian count threshold for radius $r = \SI{6}{\meter}$.}
    \parbox{\linewidth}{\footnotesize Source: \citep{seshadri_asped_2024}}
    \label{fig:cf}
\end{figure}

For Experiment~3, Figure~\ref{fig:cf} visualizes the macro accuracy for each permutation of combinations of testing and training threshold in pedestrian count, with $N_\mathrm{Thresh} \in [1,2,3,4]$. We can make the following observations: First, during testing, we see a greater proportion of correctly classified samples as the pedestrian count increases across all permutations. This is unsurprising since a larger count of pedestrian activity likely correlates to stronger and more easily detected signals. Second, while increasing the threshold during training, we find that performance generally decreases, which implies that the classifier training benefits from more difficult samples. Overall, we find optimal performance when trained with low pedestrian counts, classifying samples with high pedestrian counts (upper right box of Figure~\ref{fig:cf}). 

\subsection{Pedestrian Flow Detection}

In predicting the number of pedestrians around the four levels of radii (\SI{1}{\meter}, \SI{3}{\meter}, \SI{6}{\meter}, and \SI{9}{\meter}), the prediction accuracy is presented in Table~\ref{tab:accuracy}. The CNN demonstrated high accuracy in predicting pedestrian numbers within a \SI{1}{\meter} radius of each recorder, likely due to a prevalence of zero values in this range. Its performance was similarly strong for a \SI{3}{\meter} radius, though with a notable decrease in accuracy at recorder location 3, possibly due to its complex positioning at a 4-way intersection. The accuracy for a \SI{6}{\meter} radius remained above 90\% for most locations, even exceeding 95\% for recorders 1, 4, and 6. However, as the radius increased to \SI{9}{\meter}, accuracy dropped to around 80\% across all locations, with recorder 4 maintaining high performance, potentially due to fewer pedestrians passing by.

Recorders 1, 2, 5, and 6 saw a significant decrease in accuracy at the \SI{9}{\meter} radius, reflecting the complexity added by the larger area, as the prediction difficulty scales with the square of the radius. Despite these variations, the CNN did not overfit and maintained consistent performance on the test set.

\begin{table}
    \centering
    \caption{Prediction accuracy by target boundary size.}
    \label{tab:accuracy}
    \begin{tabular}{c}
        \includegraphics[width=0.8\linewidth]{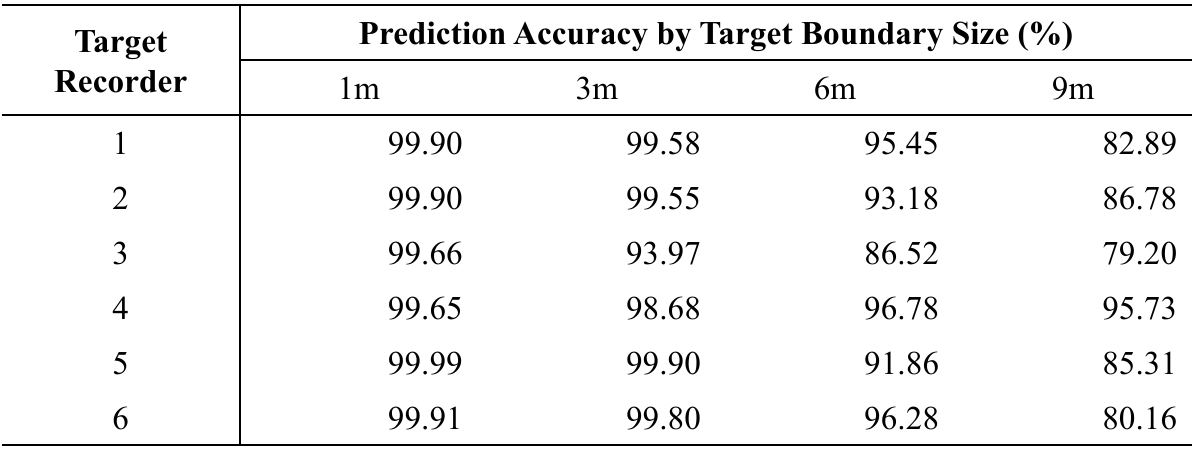}
    \end{tabular}
\end{table}

Our results indicate that, given the high accuracy for predicting the next frame based on the previous 11 frames, a sliding window method could be effective for predicting pedestrian flow over a short period. To illustrate, using the data from frames $t-11$ to $t-1$ to predict frame $t$, we can then shift the window to predict frame $t+1$ using frames $t-10$ to $t$, and so on. However, the reliability of this method is limited to a short time frame, as it doesn't account for the potential influx of new pedestrians over longer periods.

For our next steps on large-scale pedestrian flow prediction, we aim to strategically position sensors during our next data collection. We will gather comprehensive data on pedestrian movement in various directions, essential for effectively training our model to predict the inflow and outflow. Additionally, we plan to implement GCN. This approach is similar to CNNs, but it involves transforming our input data into a graph structure rather than a 2D matrix. In this graph format, sensor locations will be represented as nodes, and edges will symbolize the relative influence factors between nodes, including aspects such as topography, physical distances, and types of road segments.

\section{Discussion and Conclusion}\label{sec:discussion}
In this study, we conducted experiments to assess the efficacy of pedestrian sensing using audio sensors. 
As a result, the CONV and AST models outperformed the VGGISH model in pedestrian detection, with the most consistent accuracy observed at radii of 3 to 6 meters. The AST model showed a balanced performance across classes, and training methods effectively addressed data class imbalances. Accuracy varied across different pedestrian count thresholds. The best performance was achieved when models were trained on lower pedestrian counts and tested on higher counts, indicating better detection of stronger pedestrian signals. The results demonstrate promising potential for achieving reliable pedestrian prediction outcomes solely using audio sensors.

Moreover, we conducted a street-scale pilot experiment on short-term pedestrian flow prediction. Utilizing CNN effectively predicted pedestrian counts within varying radii, though we noted a decrease in accuracy as the radius increased. The experiment also highlights the potential of the sliding window method for short-term prediction using recent data for continuous forecasting. 

These insights lay the groundwork for our following discussion, where we aim to address challenges associated with the dataset, identify future directions for improving the reliability of audio sensing, and examine the role of pedestrian flow prediction in urban planning, with a particular focus on the contribution of audio sensing to this field. 

\subsection{Towards intricate, robust and safe audio sensing}
While our audio sensing system shows promising results, the performance of these basic algorithms is not yet comparable to video-based systems. Consequently, our immediate goal is to refine the accuracy of audio sensing. A potential limitation is that some of our audio features are trained on audio event detection, which is a more coarse-grained task than pedestrian detection. The challenge is compounded by the subtle nature of pedestrian sounds, influenced by diverse factors such as location, road surface, walking speed, and footwear, making the task particularly complex for the current audio features. Thus, we plan to develop a model specifically tailored to recognize pedestrian sounds by identifying distinctive patterns within the data.

Additionally, our dataset, primarily recorded in a campus setting, lacks the complexities of urban environments, such as vehicular noise. For the system to function effectively in a city, it must accurately detect pedestrians across various scenarios. To generalize our system to cover such scenarios, we could collect more data that includes different road conditions or use data augmentation during our training, which manually adds possible noise into our data to imitate some complex use cases.

Furthermore, privacy concerns arise from the nature of audio data, which may capture people's voices and potentially fragments of conversations containing sensitive information. To address this, we employed the Whisper speech-to-text model from OpenAI \footnote{\href{https://openai.com/research/whisper}{https://openai.com/research/whisper}, last access date Sep 5, 2023} to analyze our dataset for audible conversations. Our findings showed that in the initial dataset, conversations were either absent or not sufficiently loud and continuous for transcription. Typically, only isolated words were captured, not forming coherent sentences, likely because individuals were either consistently walking or not sufficiently close to the recorders. However, there remains a concern for scenarios where individuals decide to stop near the recorders and engage in substantive conversations. 

To mitigate the leaking of private conversations in such instances, we propose modifying the audio segments where voices are clear enough for transcription. By employing a source separation algorithm to remove voices, we can obscure any identifiable or private dialogue while maintaining the integrity of the data for our model. This approach ensures we respect individuals' privacy while benefiting from the potential benefits of audio-based pedestrian detection.

\subsection{Audio sensing for pedestrian flow prediction}

Considering the difficulties researchers often face in accessing mobile GPS data for urban planning, our publicly accessible dataset can be a crucial resource for urban crowd management and infrastructure investments. We especially anticipate it will significantly aid pedestrian flow prediction projects, ultimately leading to smarter, more effective urban design strategies.

Pedestrian flow prediction in urban settings presents significant implications for enhancing urban environments. By forecasting pedestrian flow patterns, urban planners can optimize street and road network designs to meet the dynamic needs of the city's inhabitants in each street segment at different times. Such knowledge allows for an informed categorization of streets and targeted interventions, distinguishing those bustling with activity from the underused sections. For example, areas identified as high pedestrian traffic zones can be prioritized for safety improvements, amenities, or aesthetic enhancements, while underutilized areas can be reimagined to better serve the community. Furthermore, pedestrian flow prediction has become a critical component in urban scenario planning, particularly in forecasting the impact of proposed changes in the built environment on foot traffic \citep{Sevtsuk2021}. For example, when there is a change in land use in a neighborhood or a new public infrastructure in an area, pedestrian flow prediction models could aid urban planners in assessing and simulating the impacts of these interventions. 

In this context, integrating audio sensing into pedestrian flow prediction methodologies offers an exciting avenue for enhancing the accuracy and reliability of these predictions. Unlike video surveillance, audio sensing devices maintain their effectiveness across a range of lighting conditions and can provide valuable data in areas where traditional pedestrian sensing setups might fall short. This technology's unique capabilities, including its resilience to varying light conditions and its effectiveness in densely built-up areas, position it as a complementary tool that could significantly advance our understanding and forecasting of pedestrian movement patterns, ultimately contributing to the development of more livable and responsive urban environments.

\bibliography{sn-bibliography,Zotero, Zotero_CHan}

\end{document}